\documentclass{aastex701}

\usepackage{dcolumn}
\usepackage{amsmath}
\usepackage{xcolor}

\begin{document}


\title{Possible Supermassive Dark Object Composed of Light Fermionic Gas with an Embedded Neutron Star Core}

\author[orcid=0009-0007-5786-022X, gname=Daichen, sname=Zou]{Daichen Zou}
\altaffiliation{These authors contributed equally to this work.}  
\affiliation{SDU-ANU Joint Science College, Shandong University, Weihai, China}
\email{zoudaichen289@gmail.com}

\author[orcid=0009-0000-6744-3074, gname=Xudong, sname=Wang]{Xudong Wang}
\altaffiliation{These authors contributed equally to this work.}  
\affiliation{Shandong Provincial Key Laboratory of Nuclear Science, Nuclear Energy Technology and Comprehensive Utilization, Weihai Frontier Innovation Institute of Nuclear Technology, School of Nuclear Science, Energy and Power Engineering, Shandong University, Shandong 250061, China}
\email{fakeemail2@google.com}

\author[orcid=0000-0001-7159-1897]{Bin Qi}
\altaffiliation{corresponding author}
\affiliation{Shandong Provincial Key Laboratory of Nuclear Science, Nuclear Energy Technology and Comprehensive Utilization, Weihai Frontier Innovation Institute of Nuclear Technology, School of Nuclear Science, Energy and Power Engineering, Shandong University, Shandong 250061, China}
\email[show]{bqi@sdu.edu.cn}

\begin{abstract}

The structure of dark matter admixed neutron stars (DANSs) are investigated,  adopting a non-annihilating self-interacting fermionic dark matter (DM) model,  with a particular focus on the case of the light DM particle mass \(m_D \in [10^{-10}, 1]\) GeV.  The  DANSs become DM-dominated configurations when \(m_D <10^{-1}\) GeV, where a compact neutron star core becomes embedded within an extremely large DM halo. It is found that the maximum mass of DANSs is inversely proportional to $m_{ D}$, approximately as $ 0.627 (\mathrm{GeV/} m_{\rm D})^2 ~\mathrm{M_{\odot}}$, which implies that extremely large masses can be achieved for small $m_{\rm D}$. For $m_D \sim5\times10^{-4}$ GeV, the calculated  mass and size of the DM halo can be comparable to those of supermassive black holes such as  Sgr A*. Our findings hint at a scenario where neutron stars might serve as strong gravitational seeds for such supermassive dark objects.


\end{abstract}



\section{Introduction}
Supermassive compact objects have garnered considerable attention over the past several decades \citep{Katz20, BecerraVergara21, Vincent21}. It is widely believed that these objects are black holes (BHs) \citep{Toubiana21, Vincent21}. Recently, a possible alternative to BHs has been put forward, motivated by a substantial body of research focused on Sgr A*, which posits the existence of supermassive compact objects composed of self-gravitating fermionic dark matter (DM) \citep{Bilic02, Ruffini15, Arguelles16, Gomez16, Saxton16, Arguelles18, BecerraVergara20, BecerraVergara21, Arguelles22, Pelle24}. The proposal for such a supermassive compact object is formulated within the framework of the RAR model \citep{Ruffini15, Arguelles18}.
The extended RAR model contains equilibrium solutions for the DM density distribution, exhibiting a core-halo morphology: a dense fermionic core that can mimic Sgr A* and an extended halo that accounts for the rotation curve of the Milky Way \citep{BecerraVergara20, BecerraVergara21, Arguelles22}.

As is well known, the existence of a gravitational potential well promotes the accretion of DM. Compact stars like neutron stars exert a huge gravitational effect, resulting in the accretion of DM and the generation of dark matter admixed neutron stars (DANSs). Extensive research for DANSs has been performed \citep{Panotopoulos17,Fornal18,Rezaei18,Wang19,PerezGarcia20,Husain22,Cronin23,Barbat24,Liu24,Sun24,Viklaris24,Bell25,Hajkarim25,Kumar25}, demonstrating that a significant amount of DM can indeed be accommodated within neutron stars.

Much of the research of DANS has focused on Weakly Interacting Massive Particles (WIMPs), one of the candidates of DM, which were long considered the most promising candidates for DM over the past several decades \citep{Bertone18}. Several methods are used to detect DM, including the direct detection of DM scattering signals \citep{Narain06,Amare19,Cui22,Aprile25}, the detection of new products or accompanying particles in particle accelerators \citep{Arcadi12,LeJoubioux24}, or the observation of the production of DM annihilation \citep{Feng13,Arguelles21}. These experiments are insufficient to provide definitive evidence for the existence of WIMPs. In contrast, recent studies \citep{Cox25,DeLaTorreLuque25,Fukuda25,Nagao25,Zatini26} have increasingly turned to light DM particles, which are emerging as promising alternative candidates.

On the other hand, DANSs composed of WIMPs are incapable of forming supermassive dark objects \citep{Panotopoulos17,Wang19,Barbat24}. For instance, the maximum DM mass is only about $0.27$ $\mathrm{M_{\odot}}$ and $0.027$ $\mathrm{M_{\odot}}$ for $m_D = 10$ $\mathrm{GeV}$ and $100$ $\mathrm{GeV}$, respectively \citep{Wang19}. These values are far below the mass of Sgr A*, implying that a much smaller $m_D$ would be required to achieve a larger total mass.
 
To date, very few studies have investigated the nature of DANSs for $m_D$ below $1$ $\text{GeV}$ \citep{Barbat24,Liu24,Viklaris24}. \cite{Viklaris24} predicted the existence of celestial bodies with diameters of hundreds of kilometers and masses exceeding $100$ $\mathrm{M_{\odot}}$ when $m_{D} = 0.01$ $\mathrm{GeV}$.
This motivates us to investigate whether, for even lower $m_{D}$, the DM halo within a DANS could evolve into structures analogous to the supermassive compact objects discussed earlier. We therefore turn attention to $m_{D}$ in the range $[10^{-10}, 1]$ $\mathrm{GeV}$ and study its impact on the structure of DANSs.

The structure of this article is arranged as follows. 
In Section~\ref{sec:formula}, we introduce the theoretical framework, including the two-fluid TOV equations, the equations of state (EOSs) for fermionic DM and NM.
Section~\ref{sec:structure} presents the structural properties of DANSs with light DM particles and demonstrates the emergence of DM‑dominated configurations. 
An empirical scaling relation for the maximum mass of such systems is also presented. It will be found that extremely large masses such as  Sgr A* can be achieved for small $m_{\rm D}$ .
In Section~\ref{sec:origin}, we discuss the connection between our calculated DANSs and supermassive compact DM objects as alternatives to BHs. 
Finally, Section~\ref{sec:summary} summarizes our main findings.

\section{Formula}\label{sec:formula}
\subsection{Tolman-Oppenheimer-Volkoff equation}
The compact stars which are made of DM and NM are inherently two-fluid system. If NM and DM couple essentially only through gravity, DANSs can be studied by the TOV equations for two-fluid separately \citep{Sandin09,Leung11,Tolos15,Mukhopadhyay16,Viklaris24}:
\begin{eqnarray}
\frac{dp_D}{dr} &=& -\frac{GM(r)\varepsilon_D(r)}{r^2} \left(1+\frac{p_D(r)}{\varepsilon_D(r)}\right)
\times\left(1+4\pi r^3\frac{p_D(r)+p_N(r)}{M(r)}\right)
\left(1-2G\frac{M(r)}{r}\right)^{-1},\nonumber \\
\frac{dp_N}{dr}&=&-\frac{GM(r)\varepsilon_N(r)}{r^2} \left(1+\frac{p_N(r)}{\varepsilon_N(r)}\right)
\times\left(1+4\pi r^3\frac{p_D(r)+p_N(r)}{M(r)}\right)
\left(1-2G\frac{M(r)}{r}\right)^{-1}, \nonumber \\
\frac{dM_D}{dr}&=&4\pi r^2\varepsilon_D(r),\nonumber \\
\frac{dM_N}{dr}&=&4\pi r^2\varepsilon_N(r), \nonumber \\
M(r)&=&M_D(r)+M_N(r),
\end{eqnarray}
where $M(r)$  represents the total mass at the radius $r$.  $p_D$, $p_N$, $\varepsilon_D$, $\varepsilon_N$ represent the pressure and energy density of DM and NM, respectively.

\subsection{EOS for DM}
In this paper, we use the non-self-annihilating self-interacting fermionic model to simulate DM, where the detailed formalism is given by \cite{Mukhopadhyay16}. The EOS, i.e., energy density and pressure are written as:

\begin{eqnarray}
\varepsilon&=&\frac{1}{\pi^2}\int_{0}^{k_F}k^2\sqrt{m_D^2+k^2}dk+\left[\left(\frac{1}{3\pi^2}\right)^2y^2z^6\right]\nonumber,\\
p&=&\frac{1}{3\pi^2}\int_{0}^{k_F}\frac{k^4}{\sqrt{m_D^2+k^2}}dk+\left[\left(\frac{1}{3\pi^2}\right)^2y^2z^6\right],
\label{eq2}
\end{eqnarray}
where $m_D$ is the particle mass of fermionic DM, $k$ is the momentum, $z=k_F/m_D$ is the dimensionless Fermi momentum and $y$ is the dimensionless interaction strength
parameter, which is defined as $ y = m_D/m_I$ ($m_I$ is the interaction mass scale) \citep{Narain06,Mukhopadhyay16}. The typical scale is $m_I$$\sim300$ GeV  for weak  interaction (WI) DM, and  $m_I$ is assumed to be $\sim 0.1$ GeV for strong interaction (SI) DM, according to the gauge theory of WI and SI~\citep{Narain06,Mukhopadhyay16}.

Meanwhile, we consider the fermionic DM particle mass $m_D$  as a free parameter, with values from $10^{-10}$ GeV to $1$ GeV, encompassing the light DM regime.

\subsection{EOS for NM}
For the NM in DANSs, we adopt the Relativistic Mean-Field (RMF) theory, which has achieved great success in the description of nuclear matter and finite nuclei in the past several decades~\citep{Meng06}. Meanwhile, the RMF theory has been used to study the neutron stars and obtained a lot of valuable results. 
The start of the RMF theory is an effective Lagrangian density. In the present work, we use the density dependent RMF theory where the effective Lagrangian density with baryons, mesons ($ \sigma, \omega $ and $ \rho $), and photons as degrees of freedom for nuclear matter. In this work, we choose DDME2 parameter set \citep{Lalazissis05}.

\section{The structure of DANS with light DM}\label{sec:structure}
\begin{figure}[h] 
    \centering
    \includegraphics[width=0.5\textwidth]{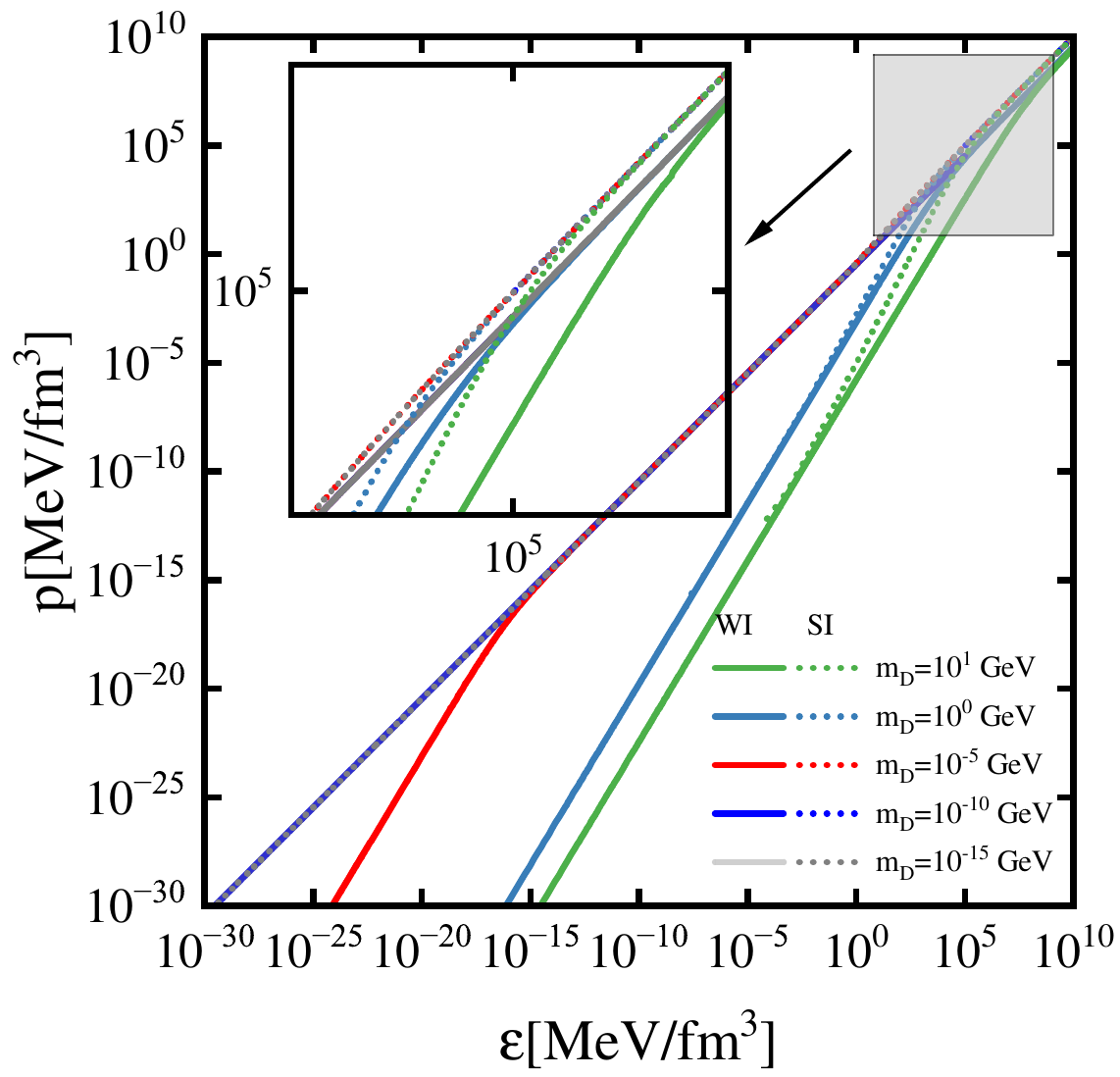} 
    \caption{The EOSs for the DM for SI  (solid lines) and WI (dotted lines), namely, the pressure as a
    function of energy density with $m_D = 10^1, 10^{0}, 10^{-5}, 10^{-10}, 10^{-15}$ GeV. Different colors represent different $m_D$.}
    \label{fig:fig1}
\end{figure}

Fig.~\ref{fig:fig1} depicts the EOSs for DM for SI  and WI, namely, the pressure as a function of energy density with $m_D = 10^1, 10^{0}, 10^{-5}, 10^{-10}, 10^{-15}$ GeV.

When $\varepsilon$ increases, all EOS curves will gradually merge into the limiting curve corresponding to the causal limit $\rho =\varepsilon$. This rule holds for both SI and WI. 
The density at which the EOS curve approaches the causality limit depends on the values of both $m_{D}$ and $m_{I}$. As $m_{D}$ decreases, the density at which it reaches the causality limit becomes lower. 
Beyond the causal limit, the  $p$ and  $\varepsilon$ expressed in Eq.~(\ref{eq2}) are dominated by the interaction term $\left( 1/{3\pi^{2}} \right)^{2} y^{2} z^{6}$, resulting in $p\approx\varepsilon$.

Prior to the causal limit, the EOSs for different $m_{D}$ exhibit a common slope of $1/3$ under the same interaction, as manifested by the dominance of the Fermi degeneracy pressure: $p \approx \varepsilon/3$, with $\varepsilon \sim (1/{\pi^2}) \int_{0}^{k_F} k^{3}  dk$. It should be pointed out that the deviation at low densities of the EOS compared with the slope of high densities is crucial for the convergence of the solution of the two-fluid TOV equation.

\begin{figure}[h] 
    \centering
    \includegraphics[width=1.0\textwidth]{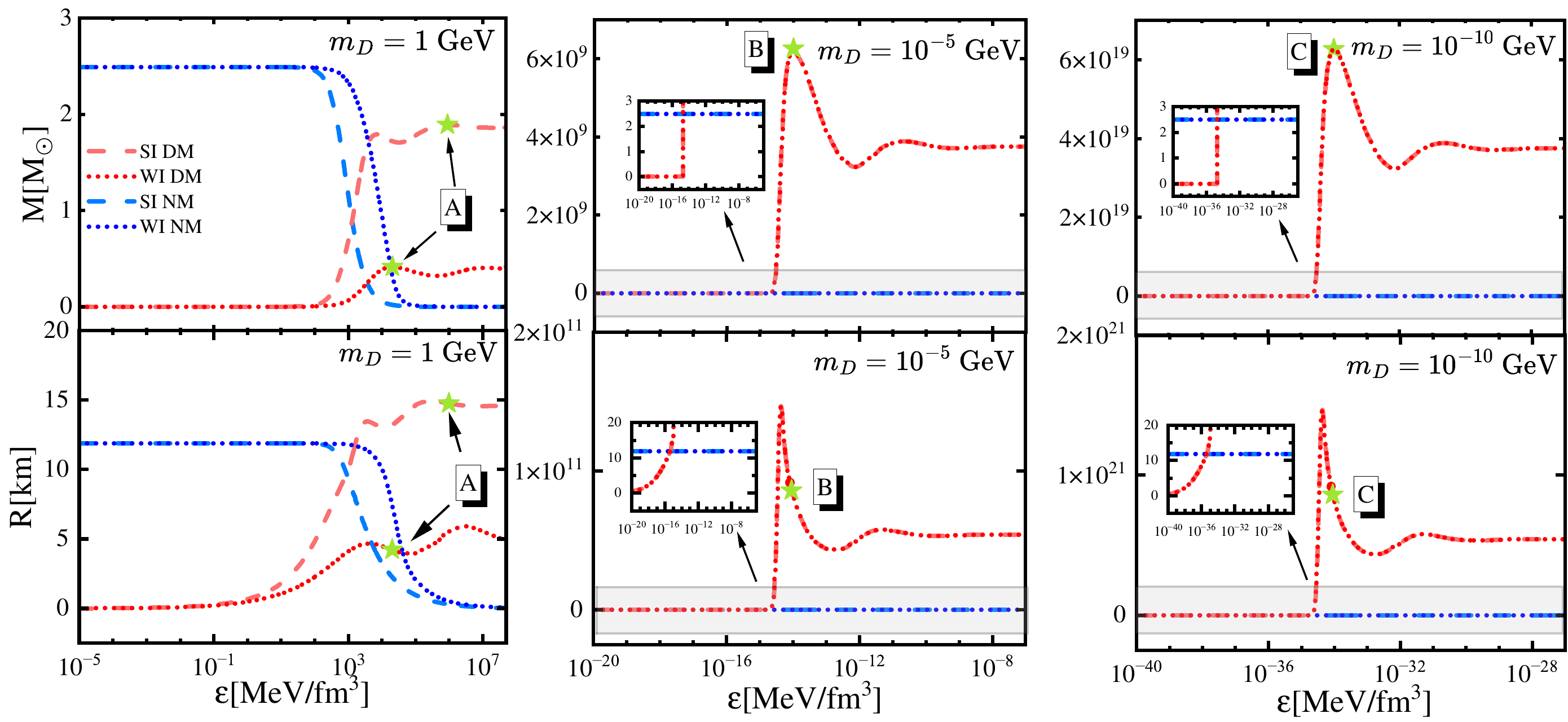} 
    \caption{The calculated mass and radius of DM and NM in DANSs as a function of the central energy density of DM $\varepsilon$ under SI and WI with $m_D = 1, 10^{-5}, 10^{-10}$ GeV. DM masses (the upper panels) and radii (the lower panels) for SI and WI are indicated by light red dashed and red dotted lines, respectively; NM masses (the upper panels) and radii (the lower panels) for SI and WI are shown as light blue dashed and blue dotted lines, respectively. The EOS of NM is from RMF with DDME2 parameter set, and the central energy density of NM is fixed at 1050 $\mathrm{MeV/fm^{3}}$.  The inset represents a zoomed-in view of the gray shaded region. And ``A, B, C'' are denoted as the cases where the $M_D$ attains its maximum.}
    \label{fig:fig-2}
\end{figure}

 Fig.~\ref{fig:fig-2} shows the calculated mass and radii of DM and NM in DANSs as a function of the central energy density of DM $\varepsilon$ under SI and WI with $m_D =  1,  10^{-5}, 10^{-10}$ GeV. DM masses $M_{D}$ and radii $R_D$ for SI and WI are indicated by light red dashed and red dotted lines, respectively, NM masses $M_{N}$ and radii $R_N$ for SI and WI are shown as light blue dashed and blue dotted lines, respectively. The EOS of NM is from RMF with DDME2 parameter set, and the central energy density of NM is fixed at 1050 $\mathrm{MeV/fm^{3}}$. The mass of a pure neutron star without DM achieves its maximum when the central energy density of NM reaches 1050 $\mathrm{MeV/fm^{3}}$.

When $m_D =  1$ GeV, at low DM central energy densities $\varepsilon$, the presence of DM exerts negligible influence on the $M_N$  which remains stable at $2.48$ $\mathrm{M_{\odot}}$. Increasing $\varepsilon$ leads to an increase in $M_D$, accompanied by a rapid decline of $M_N$ to negligible levels. For SI and WI DM, we can see that the maximum mass of DM reaches 1.89 $\mathrm{M_{\odot}}$ and 0.41 $\mathrm{M_{\odot}}$, respectively, which can affect the mass of DANSs apparently. 
It is observed that for $m_D = 1$ GeV, an increase in the amount of DM leads to a decrease in the mass of the DANSs.

When $m_D =  10^{-5}, 10^{-10}$ GeV, the results of the TOV equation for WI and SI are almost identical with each other, as shown in Fig.~\ref{fig:fig-2}. The $M$-$\varepsilon$ curve of DM shares a similar profile where the DM mass first rises to a peak and then declines with increasing $\varepsilon$, eventually stabilizing after a series of oscillations. At fixed $\varepsilon$, a decrease in $m_D$ leads to a dramatic increase in $M_D$, while $M_N$ remains constant at 2.48 $\mathrm{M}_{\odot}$. Notably, $M_D$ can attain remarkably high values, reaching up to 
$\sim 10^{9}\,\mathrm{M}_{\odot}$ and $\sim 10^{19}\,\mathrm{M}_{\odot}$, respectively.  ``A, B, C'' are denoted as the cases where the $M_D$ attains its maximum.

\begin{figure}[h!] 
    \centering

         \includegraphics[width=0.8\textwidth]{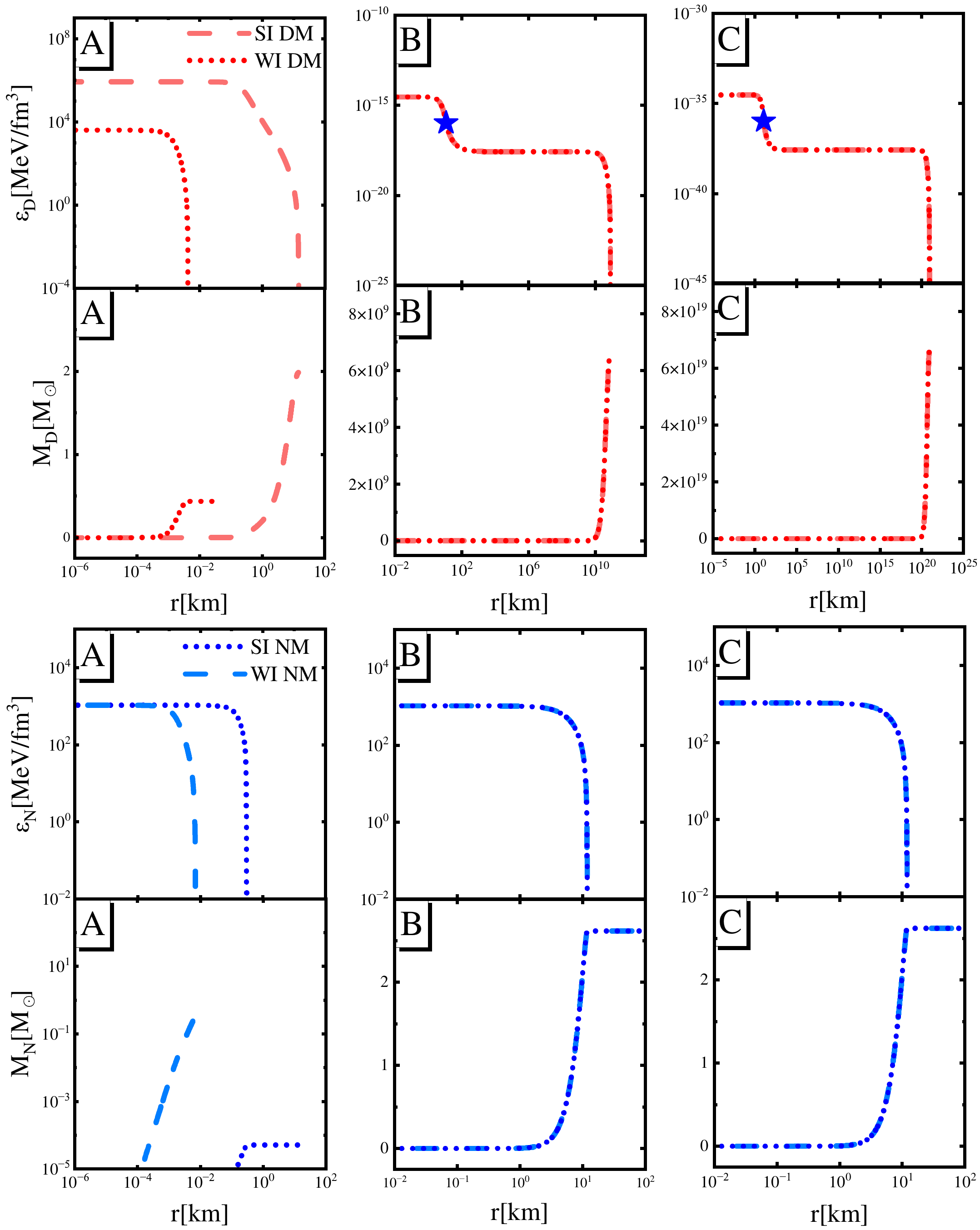} 
    \caption{The energy densities $\varepsilon$ and masses as functions of $r$ (distance to the center of stars) for the DM and NM corresponding to three specific cases ``A, B, C'' when DM mass achieves its maximum. DM for SI and WI are indicated by light red dashed and red dotted lines, respectively; NM for SI and WI are shown as light blue dashed and blue dotted lines, respectively.}
    \label{fig:fig-3}
\end{figure}

The \(R\)-\(\varepsilon\) curves display characteristics similar to those of the corresponding \(M\)-\(\varepsilon\) curves.  The radius of the DM halo (core) at the maximum mass is not the maximum radius of DM halo; its value is slightly lower than the maximum but remains within the same order of magnitude.  For instance, when $m_D = 10^{-10}$ GeV, the DM radius at the maximum mass is about $8.58\times10^{20}$ km, while the maximum DM radius is approximately $1.5\times10^{21}$ km.

In Fig.~\ref{fig:fig-3}, we choose three typical cases which are the same as those denoted by ``A, B, C'' in Fig.~\ref{fig:fig-2} to show their energy densities $\varepsilon$ and masses as functions of $r$ (distance to the center of stars) when the mass of DM achieves maximum.

Interestingly,  case (B) and case (C) reveal a kind of supermassive dark object when $m_D$ lies within the sub-GeV mass range. The DM mass and radius take values of $6.58\times10^{9}$ $\mathrm{M_{\odot}}$ and $7.86\times10^{10}$ km for $m_D = 10^{-5}$ GeV, 
with $6.58\times10^{19}$ $\mathrm{M_{\odot}}$ and $7.86\times10^{20}$ km for $m_D = 10^{-10}$ GeV, respectively. For both cases, the $M$--$R$ profile of NM is precisely the same, with $M_{N} = 2.48$ $\mathrm{M_{\odot}}$ and $R_{N} = 11.7$ km. This implies that a stable neutron star serves as the core embedded within a supermassive DM halo.
The blue star in Fig.~\ref{fig:fig-3} denotes the radius of the neutron star core, which lies in the region where the DM density undergoes an abrupt change. This region corresponds to the boundary separating the two-fluid TOV and the single-fluid TOV regimes.

 \begin{figure}[h] 
    \centering
    \includegraphics[width=0.6\textwidth]{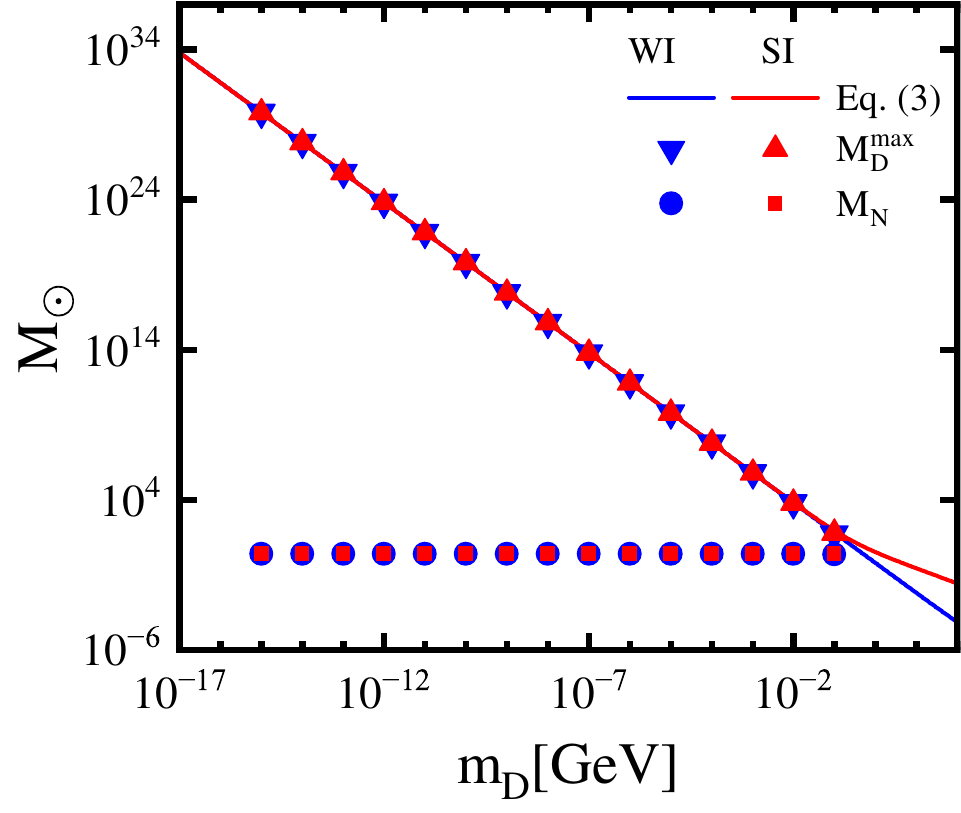} 
    \caption{The maximum DM mass (Blue inverted triangles and red upright triangles) and the NM mass (Blue circles and red squares) versus $m_D$ for DANSs for SI and WI  cases, respectively. The red line and the blue line correspond to the suggested relationship of SI and WI in Eq.~(\ref{eq:eq16}).}
    \label{fig:fig4}
\end{figure}
\cite{Narain06} demonstrated there exists a single-dependent relationship in pure DM stars between the maximum DM mass $M_{D}^{max}$ and particle mass $m_{D}$ by deriving the scaling solution of self-interacting Fermi gas of the TOV equation, which can be written as
 \begin{equation}
M_{D}^{max} = (0.269\frac{m_{D} }{m_{I} }+0.627 )(\frac{1 \mathrm{GeV}}{m_{D}} ) ^{2}\mathrm{M_{\odot }} .
\label{eq:eq16}
\end{equation}

To examine whether the relationship of Eq.~(\ref{eq:eq16}) still holds for DM halo of DANSs for smaller $m_D$, the maximum DM mass $M_{D}^{max}$ and its corresponding NM mass $M_{N}$ as a function of $m_{D}$ are given in the Fig.~\ref{fig:fig4}, in which two different interaction strengths are adopted. It can be seen that the relationship of Eq.~(\ref{eq:eq16}) is still valid for DM halo of DANSs when $m_D\in[10^{-15},10^{-1}]$ GeV.

The curves of Eq.~(\ref{eq:eq16}) under different interaction strength tend to be consistent with $m_{D}$ decreasing. While the curve of stronger interaction is the first to deflect towards the large mass as $m_{D}$ grows. 

The smaller the $m_D$ ($m_D /m_I \ll 1$), the more DM that a DANS can contain, and this relationship exhibits a power-law scaling: 
 \begin{equation}
 M^{max}_{DANS} = 0.627  (\frac{1 \mathrm{GeV}}{m_{D}} ) ^{2}\mathrm{M_{\odot }} , 
  \end{equation}
which implies that extremely large masses can be achieved for small $m_{\rm D}$. For instance, when $m_D = 10^{-5}\ \mathrm{GeV}$, the resulting DANS can attain a mass of $\sim 10^{9}\ \mathrm{M_{\odot}}$.

As $m_D$ decreases, $M_N$ does not change.
The structure transitions from a mixed core to a DM-dominated configuration, where a compact neutron star core becomes embedded within an extensive DM halo.

\section{SGR A* as a possible DANS}\label{sec:origin}

When $m_D$ is sufficiently light,  the NM core is tightly bound at the center of the DM gravitational potential well, much like a compact seed embedded within it. In the case of light DM, a DANS is essentially an extremely enormous DM celestial body, merely embedded with an NM core. The presence of a neutron star might be favorable for the formation of such an object,  since the neutron star itself  can serve as a strong gravitational source to accrete ambient DM, thereby facilitating the formation of such a supermassive configuration.

\begin{figure}[h!] 
    \centering
         \includegraphics[width=0.6\textwidth]{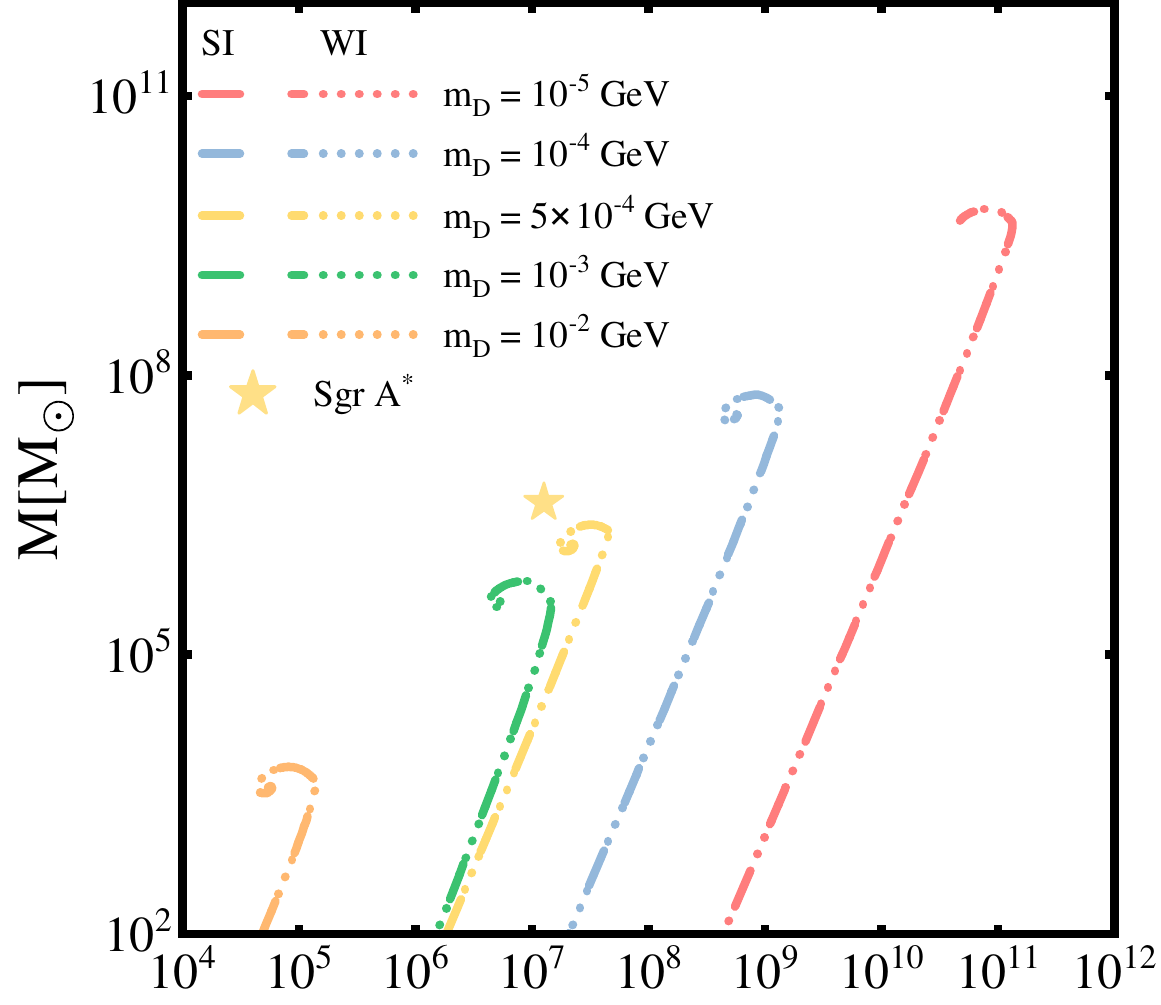} 
    \caption{The mass of DM as a function of the radius $R$ of the supermassive DM celestial bodies with different $m_D$ under SI and WI. The lines of different colors depict the $M$-$R$ curves at various  $m_D$. The particle mass $m_D$ of fermionic DM varies from $10^{-5}$ GeV to  $10^{-2}$ GeV. The golden star in the figure represents Sgr A$^{*}$.}
    \label{fig:fig-new}
\end{figure}

A recently proposed alternative  to BHs (including Sgr A*) is supermassive compact fermionic DM balls, with mass and size comparable to those of BHs~\citep{Bilic02, Ruffini15, Arguelles16, Gomez16, Saxton16, Arguelles18, BecerraVergara20, BecerraVergara21, Arguelles22, Pelle24}. 
Fig.~\ref{fig:fig-new} presents the DM mass  as a function of the radius of the supermassive DM celestial bodies with different $m_D$ under SI and WI. The solid and dashed lines of different colors depict the $M$-$R$ curves at various  $m_D$ under SI and WI. The particle mass $m_D$ of fermionic DM varies from $10^{-5}$ GeV to  $10^{-2}$ GeV. The golden star in the figure marks the total mass and Schwarzschild radius of Sgr A$^{*}$. In Fig.~\ref{fig:fig-new}, the mass and radius of the DM halo exhibit a strong dependence on $m_D$. For DM particle masses \(m_D \sim 5\times10^{-4}\) GeV (\(m_D \sim 500\) keV ), the resulting configurations closely resemble those of Sgr A\(^{*}\), thereby offering possible clues for constraining \(m_D\).  This value of $m_D$ is quite close to the DM particle mass (48 - 345\,{keV}) suggested by \cite{Arguelles18} for the BH alternative scenario of pure DM objects.
The obvious distinction between the present study and the previous work \citep{BecerraVergara21,Arguelles22} is that the proposed supermassive DM object contains a compact neutron star core at its center. 
This feature suggests one possible formation channel: the neutron star core, possibly acting as a strong gravitational seed, might accrete surrounding DM, which could then facilitate the formation of such an object.

This work  focuses on calculating the static equilibrium structures and maximum mass limits of DANSs for light fermionic DM. Our results demonstrate that, from a theoretical perspective, static configurations with supermassive DM halos are permissible within the light DM mass.  Several theoretical models have been proposed to describe the accretion process of DM by neutron stars \citep{Gould87,Li12,PerezGarcia12,Guver14}, inconsistencies among their models indicate that the underlying mechanisms require further clarification. The question of whether such equilibrium states of supermassive DANSs could be reached via dynamical accretion processes, remains open for future study.

The structure proposed in this work---a central dense neutron star core surrounded by a supermassive extended  DM halo, whose morphology qualitatively matches the fermionic DM structures of the RAR model \citep{Arguelles18,BecerraVergara20,BecerraVergara21,Arguelles22,Pelle24}. For masses \(48-345\,\mathrm{keV}\), the same DM core–halo distribution simultaneously fits the Milky Way rotation curve from sub‑pc to kpc scales \citep{Arguelles18}, offering a cross-scale consistency check. At horizon scales, general relativistic ray‑tracing shows that cores with $m_D$ $\in [300, 378]$ keV generate a central brightness depression surrounded by a ring-like structure when illuminated by a thin accretion disc \citep{Pelle24}.   Future very-long-baseline interferometry, may detect the presence or absence of photon rings, then to test which of the hypotheses — black holes or dark matter objects — is reasonable~\citep{Arguelles21}.

\section{Summary}\label{sec:summary}
In this work, we investigate the structure of DANSs composed of fermionic DM and NM, with a particular focus on the light DM particle mass \(m_D \in [10^{-10}, 1]\) GeV. Employing the two‑fluid TOV equations, DM is modeled as non‑annihilating self‑interacting Fermi gas, while NM is described within the relativistic mean‑field theory.

It is found that the maximum mass of DANSs follows an inverse‑square scaling with $m_D$, approximately  $0.627 (\mathrm{GeV/} m_{\rm D})^2 ~\mathrm{M_{\odot}}$. This relationship implies that extremely large DM masses can be achieved for small \(m_{\mathrm{D}}\). As $m_D$ decreases,  the DANSs become a DM‑dominated configuration, where a compact neutron star core remains essentially unchanged and is embedded within an extensive DM halo.  For $m_D  \sim 5\times10^{-4}$  GeV, the resulting object matches the mass and radius of Sgr~A* remarkably well, providing a direct link to the recently proposed alternative to the BH paradigm.   Our findings hint at a scenario where neutron stars might serve as strong gravitational seeds for such supermassive dark objects.

\section*{Acknowledgments}
This work is partly supported by the National Natural Science Foundation of China (No.~12475123, No.~12225504, No.~12321005).

\begin{contribution}

Daichen Zou and Xudong Wang contributed equally to this work.


\end{contribution}

%



\bibliography{sample701}{}
\bibliographystyle{aasjournalv7}

\end{document}